\documentclass{article}[12pt]

\usepackage[margin=1in]{geometry}
\usepackage{amsmath, amsfonts, amssymb}
\usepackage{graphicx}
\usepackage{booktabs}
\usepackage{natbib}
\usepackage{caption}
\usepackage{subcaption}
\usepackage{multirow}
\usepackage{array}
\usepackage{longtable}
\usepackage{xcolor} 
\usepackage{siunitx} 
\usepackage{algorithm} 
\usepackage{algpseudocode} 
\usepackage{hyperref} 

\newcommand{\vect}[1]{\mathbf{#1}} 
\newcommand{\mat}[1]{\mathbf{#1}} 

\title{Sampled Grid Pairwise Likelihood (SG-PL): \\An Efficient Approach for Spatial Regression on Large Data}
\author{Giuseppe Arbia\thanks{Università Cattolica del Sacro Cuore, Rome, Italy} \and Vincenzo Nardelli\footnotemark[1] \and Niccolò Salvini\footnotemark[1]}
\date{}

\begin{document}

\maketitle

\begin{abstract}
\noindent Estimating spatial regression models on large, irregularly structured datasets poses significant computational hurdles. While Pairwise Likelihood (PL) methods offer a pathway to simplify these estimations, the efficient selection of informative observation pairs remains a critical challenge, particularly as data volume and complexity grow. This paper introduces the Sampled Grid Pairwise Likelihood (SG-PL) method, a novel approach that employs a grid-based sampling strategy to strategically select observation pairs. Simulation studies demonstrate SG-PL's principal advantage: a dramatic reduction in computational time—often by orders of magnitude—when compared to benchmark methods. This substantial acceleration is achieved with a manageable trade-off in statistical efficiency. An empirical application further validates SG-PL's practical utility. Consequently, SG-PL emerges as a highly scalable and effective tool for spatial analysis on very large datasets, offering a compelling balance where substantial gains in computational feasibility are realized for a limited cost in statistical precision, a trade-off that increasingly favors SG-PL with larger N.\end{abstract}

\noindent\textbf{Keywords:} Pairwise Likelihood, Spatial Regression, Large Datasets, Irregular Grids, Computational Efficiency.

\newpage

\section{Introduction}
The proliferation of geospatial technologies, including remote sensing, social media, Internet of Things (IoT) devices, and detailed administrative records, has led to an unprecedented availability of large spatial datasets. These datasets offer rich opportunities for understanding complex spatial phenomena. However, their size poses significant computational challenges for traditional spatial econometric models, such as Spatial Error Models (SEM). Exact likelihood methods for these models often involve the computation of determinants or inversions of large $N \times N$ spatial weight matrices or variance-covariance matrices, leading to computational complexities of $O(N^3)$ or $O(N^2)$ in the best cases. When $N$ (the number of observations) reaches tens of thousands to millions, such methods become computationally prohibitive.

Pairwise Likelihood (PL), a specific form of composite likelihood \citep{Lindsay1988}, offers a computationally attractive alternative. The core idea of PL is to simplify the full joint likelihood by summing or multiplying the likelihoods of smaller subsets of data, typically pairs of observations. \citep{Arbia2014} proposed an extension of PL on regular-gridded dataset on the estimation of Spatial Error Model (SEM). While this approach offers theoretical advantages, its practical application, especially on irregular grids (i.e., point-referenced data not aligned on a regular lattice), requires efficient strategies for selecting observation pairs. The goal is to select pairs that are informative about the underlying spatial process while ensuring that the computational burden of pair selection itself does not negate the benefits of PL. Recent efforts, such as the KD-Tree based approach for irregular grids by \citet{ArbiaSalvini2024}, aim to address this challenge by providing structured ways to identify neighboring or relevant pairs.

This paper introduces the "Sampled Grid Pairwise Likelihood (SG-PL)" method as its main contribution. SG-PL is designed to be a fast and straightforward way to pick pairs of observations for Pairwise Likelihood estimation, especially when dealing with large and irregularly spread out spatial data where usual methods struggle. Its core idea involves laying a regular, discrete global grid system (like H3 hexagons) over the data, identifying cells with a sufficient number of observations, and then iteratively choosing a subset of these cells for pair sampling. This selection process ensures spatial separation between the chosen cells by excluding neighbors of an already selected cell, which aims to approximate independence between pairs from different cells. Finally, one pair of observations is sampled from within each of these selected, isolated grid cells.
We hypothesize that SG-PL offers significant computational speed-up due to its structured sampling approach, which avoids all-pairs distance calculations, and its ease of implementation. Furthermore, we anticipate that it can yield reasonable statistical performance for key parameters, especially for detecting spatial dependence.

The remainder of this paper is structured as follows: Section \ref{sec:background} reviews the fundamentals of Pairwise Likelihood for spatial regression. Section \ref{sec:sgpl_method} details the proposed SG-PL methodology. Section \ref{sec:simulation} presents the design and results of a comprehensive Monte Carlo simulation study comparing SG-PL to a benchmark method. Section \ref{sec:empirical} discusses an empirical application using a large dataset of house sales. Finally, Section \ref{sec:conclusion} concludes the paper and outlines directions for future research.

\section{Background: Pairwise Likelihood for Spatial Regression}
\label{sec:background}
Pairwise Likelihood (PL) methods, as a class of composite likelihoods \citep{Varin2011}, have gained attention for their ability to handle complex dependencies in large datasets where full likelihood estimation is intractable. The fundamental idea is to construct a pseudo-likelihood function based on pairs of observations, rather than the full set of $N$ observations.

Consider a basic spatial regression model for an observation $i$:
\begin{equation} \label{eq:spatial_reg_model_scalar}
    y_i = x_i\beta + \epsilon_i, \quad i=1, \dots, N
\end{equation}
where $y_i$ is the dependent variable, $x_i$ is a scalar explanatory variable, $\beta$ is a scalar coefficient, and $\epsilon_i$ is the error term. The framework can be extended to multiple regressors, where $\vect{x}_i$ would be a vector of explanatory variables and $\beta$ a vector of coefficients, leading to matrix expressions for the estimator $\hat{\beta}_{PL}$. For clarity in presenting the iterative estimators derived from \citet{Arbia2014}, we focus on the single regressor case here. The parameter vector for this PL model is $\theta = (\beta, \sigma^2, \lambda)'$.

The Pairwise Likelihood is formed by the product of bivariate likelihoods for a selected set of $q$ pairs of observations, $S_p = \{(i_k, l_k)\}_{k=1}^q$:
\begin{equation} \label{eq:pl_product_form}
    L_p(\theta; \vect{y}, \vect{x}) = \prod_{(i,l) \in S_p} L_{il}(\theta; y_i, y_l, x_i, x_l)
\end{equation}
where $L_{il}(\theta)$ is the bivariate joint probability distribution for the pair $(i,l)$.

Following \citet{Arbia2014}, in a spatial error context, it is assumed that for any selected pair of locations $(i,l) \in S_p$ (often where $l$ is considered a "neighbor" of $i$, denoted $l \in \mathcal{N}(i)$), the corresponding error terms $(\epsilon_i, \epsilon_l)$ follow a bivariate Gaussian distribution:
\begin{equation} \label{eq:bvn_errors}
    \begin{pmatrix} \epsilon_i \\ \epsilon_l \end{pmatrix} \sim \text{BVN} \left( \begin{pmatrix} 0 \\ 0 \end{pmatrix}, \sigma^2 \mat{\Omega} \right)
\end{equation}
where $\sigma^2$ is the marginal variance of the errors, and the $2 \times 2$ correlation matrix is:
\begin{equation} \label{eq:omega_matrix}
    \mat{\Omega} = \begin{pmatrix} 1 & \lambda \\ \lambda & 1 \end{pmatrix}.
\end{equation}
Here, $\lambda \in (-1, +1)$ represents the pairwise spatial correlation between errors at locations $i$ and $l$.

The bivariate probability density function for the errors of a pair $(i,l)$ is then:
\begin{equation} \label{eq:bvn_density_errors}
    f(\epsilon_i, \epsilon_l \mid \sigma^2, \lambda) = \frac{1}{2 \pi \sigma^2 \sqrt{1-\lambda^2}} \exp \left\{-\frac{1}{2 \sigma^2\left(1-\lambda^2\right)}\left[\epsilon_i^2-2 \lambda \epsilon_i \epsilon_l+\epsilon_l^2\right]\right\}.
\end{equation}
The pairwise log-likelihood function, which is typically maximized, is the sum of the logarithms of these bivariate densities:
\begin{equation} \label{eq:pairwise_loglik_sum}
    l_p(\theta; \vect{y}, \vect{x}) = \sum_{(i,l) \in S_p} \log f(\epsilon_i, \epsilon_l \mid \sigma^2, \lambda).
\end{equation}
Substituting $\epsilon_i = y_i - x_i\beta$ and $\epsilon_l = y_l - x_l\beta$, the contribution of a single pair $(i,l)$ to this sum is:
\begin{equation} \label{eq:pairwise_loglik_contrib_single}
\begin{split}
    \log f(y_i, y_l \mid x_i, x_l, \theta) = &-\log(2\pi) - \log(\sigma^2) - \frac{1}{2}\log(1-\lambda^2) \\
    &- \frac{1}{2\sigma^2(1-\lambda^2)} \left( (y_i - x_i\beta)^2 - 2\lambda(y_i - x_i\beta)(y_l - x_l\beta) + (y_l - x_l\beta)^2 \right).
\end{split}
\end{equation}
\citet{Arbia2014} demonstrates that PL estimators can be obtained by solving a system of equations derived from the first-order conditions. This involves the definition of six sufficient statistics, calculated over the $q = |S_p|$ selected pairs:
\begin{align}
    \alpha_1 &= \sum_{(i,l) \in S_p} (x_i^2 + x_l^2) \label{eq:alpha1} \\
    \alpha_2 &= \sum_{(i,l) \in S_p} (y_i^2 + y_l^2) \label{eq:alpha2} \\
    \alpha_3 &= \sum_{(i,l) \in S_p} (x_i y_i + x_l y_l) \label{eq:alpha3} \\
    \alpha_4 &= \sum_{(i,l) \in S_p} (x_i y_l + x_l y_i) \label{eq:alpha4} \\
    \alpha_5 &= \sum_{(i,l) \in S_p} (x_i x_l) \label{eq:alpha5} \\
    \alpha_6 &= \sum_{(i,l) \in S_p} (y_i y_l) \label{eq:alpha6}
\end{align}
The PL estimators $\hat{\beta}_{PL}$, $\hat{\sigma}^2_{PL}$, and $\hat{\lambda}_{PL}$ are then found by iteratively solving the following system of equations until convergence:
\begin{equation} \label{eq:beta_hat_pl_new}
    \hat{\beta}_{PL} = \frac{\alpha_3 - \hat{\lambda}_{PL} \alpha_4}{\alpha_1 - 2 \hat{\lambda}_{PL} \alpha_5}
\end{equation}
\begin{equation} \label{eq:sigma_sq_hat_pl_new}
    \hat{\sigma}_{PL}^2 = \frac{\alpha_2 + \hat{\beta}_{PL}^2 \alpha_1 - 2 \hat{\beta}_{PL} \alpha_3 - 2 \hat{\lambda}_{PL} \alpha_6 - 2 \hat{\lambda}_{PL} \hat{\beta}_{PL}^2 \alpha_5 + 2 \hat{\lambda}_{PL} \hat{\beta}_{PL} \alpha_4}{2q(1-\hat{\lambda}_{PL}^2)}
\end{equation}
\begin{equation} \label{eq:lambda_hat_pl_new}
    \hat{\lambda}_{PL} = \frac{\alpha_6 - \hat{\beta}_{PL} \alpha_4 + \hat{\beta}_{PL}^2 \alpha_5}{q \hat{\sigma}_{PL}^2}
\end{equation}

This PL approach offers significant advantages, including the avoidance of specifying or manipulating a full $N \times N$ spatial weights matrix or covariance matrix for the PL estimation part. The iterative solution, utilizing closed-form expressions for each parameter at each step, also contributes to computational efficiency.

However, a critical challenge in applying PL, especially to large and irregularly spaced datasets, lies in the selection of pairs $S_p$. Ideally, pairs should be "internally correlated but mutually stochastically independent" \citep{Arbia2014}. This means that while $\epsilon_i$ and $\epsilon_l$ within a pair $(i,l)$ are correlated, different pairs, say $(i,l)$ and $(j,k)$, should be approximately independent. Defining appropriate neighborhoods and ensuring selected pairs are sufficiently far apart can be computationally intensive and has motivated methods like the KD-Tree PL \citep{ArbiaSalvini2024} and the SG-PL method proposed in this paper.

While methods like the KD-Tree PL \citep{ArbiaSalvini2024} provide structured ways to identify neighboring pairs, thereby ensuring correlation within each pair $(i,l)$, they may not explicitly enforce spatial separation between different selected pairs, say $(i,l)$ and $(j,k)$. This lack of enforced distance could mean that the crucial assumption of different pairs being mutually stochastically independent is not fully met. The SG-PL method, detailed in Section \ref{sec:sgpl_method}, directly addresses this challenge through its iterative cell selection and isolation mechanism. By design, SG-PL aims to ensure that pairs are sampled from spatially distinct regions, thereby better approximating the independence condition between pairs from different cells, which is vital for the properties of the PL estimator.

\section{The Proposed Sampled Grid Pairwise Likelihood (SG-PL) Method}
\label{sec:sgpl_method}

The Sampled Grid Pairwise Likelihood (SG-PL) method is designed to provide a computationally efficient and scalable approach for selecting pairs for PL estimation, particularly for large datasets of irregularly spaced point observations. The core idea of SG-PL is to leverage a regular grid overlay to partition the irregular point pattern. This discretization of space serves multiple purposes. Firstly, it provides a structured way to manage and query spatial data, even if the underlying data points are irregularly distributed. Secondly, it facilitates a computationally efficient sampling approach by limiting search operations to within grid cells and their neighborhoods. Thirdly, and crucially, it allows for a systematic way to ensure spatial separation between selected pairs, which is important for approximating the independence assumption between pairs drawn from different regions. For this grid system, we use the H3 discrete global grid system \citep{UberH3} as an example, which tessellates the Earth's surface with hexagonal cells of roughly equal area at various resolutions. Hexagons offer advantages in neighborhood definition compared to square grids.

\subsection{SG-PL Algorithm}
The SG-PL algorithm has several user-defined parameters that can influence its performance:
\begin{itemize}
    \item $R_{H3}$ (H3 resolution): This determines the size of the hexagonal grid cells. Smaller cells (higher $R_{H3}$) mean that pairs are sampled from more localized areas, potentially capturing finer-scale spatial correlation. However, this might lead to fewer points per cell, reducing the number of candidate cells. Larger cells (lower $R_{H3}$) average over wider areas and are more likely to contain multiple points but might obscure very local patterns.
    \item $N_{\min,H3}$ (minimum points per cell): This threshold ensures that selected cells have sufficient data to form at least one pair (typically $N_{\min,H3}=2$).
    \item $k_{\text{ring}}$ (isolation buffer): This parameter controls the degree of spatial separation between the cells selected for pair sampling. A larger $k_{\text{ring}}$ increases the separation, strengthening the approximation of independence between pairs from different cells, but it also reduces the number of cells that can be selected from a finite study area, potentially limiting $q$.
    \item $q_{\text{target}}$ (target number of pairs/cells): This determines the total number of pairs used in the PL estimation. A larger $q_{\text{target}}$ incorporates more information, which can lead to more stable and statistically efficient estimates, but at the cost of slightly increased computation for the PL estimation itself (though still much less than full $N$).
\end{itemize}
The choice of these parameters may involve some exploratory analysis or be guided by the specific characteristics of the dataset and the spatial scale of interest.

The SG-PL algorithm proceeds as detailed in Algorithm \ref{alg:sgpl}.

\begin{algorithm}
\caption{Sampled Grid Pairwise Likelihood (SG-PL) Algorithm}
\label{alg:sgpl}
\begin{algorithmic}[1]
\State \textbf{Input:} Spatial data points $s_1, \dots, s_N$; H3 resolution $R_{H3}$; min points per cell $N_{\min,H3}$; isolation k-ring $k_{\text{ring}}$; target number of pairs $q_{\text{target}}$.
\State \textbf{Output:} Set of $q$ selected pairs $S_p$.

\Procedure{SG-PL}{$s_1, \dots, s_N, R_{H3}, N_{\min,H3}, k_{\text{ring}}, q_{\text{target}}$}
    \State \textbf{Grid Overlay:}
    \State Assign each data point $s_j$ to its H3 grid cell $h_k$ at resolution $R_{H3}$. Let $P_k$ be the set of points in cell $h_k$.
    
    \State \textbf{Candidate Cell Identification:}
    \State Initialize candidate cell set $C \leftarrow \emptyset$.
    \ForAll{H3 cells $h_k$ covering the study area}
        \If{$|P_k| \geq N_{\min,H3}$}
            \State Add $h_k$ to $C$.
        \EndIf
    \EndFor
    
    \State \textbf{Iterative Cell Selection and Isolation:}
    \State Initialize selected cell set $C_{\text{sel}} \leftarrow \emptyset$.
    \State Initialize available candidate cells $C_{\text{avail}} \leftarrow C$.
    \While{$|C_{\text{sel}}| < q_{\text{target}}$ \textbf{and} $C_{\text{avail}} \neq \emptyset$}
        \State Randomly select a cell $h_{\text{chosen}}$ from $C_{\text{avail}}$.
        \State Add $h_{\text{chosen}}$ to $C_{\text{sel}}$.
        \State Identify $\mathcal{N}_{k_{\text{ring}}}(h_{\text{chosen}})$: the $k_{\text{ring}}$-order neighborhood of $h_{\text{chosen}}$ (including $h_{\text{chosen}}$).
        \State $C_{\text{avail}} \leftarrow C_{\text{avail}} \setminus \mathcal{N}_{k_{\text{ring}}}(h_{\text{chosen}})$.
    \EndWhile
    
    \State \textbf{Pair Sampling:}
    \State Initialize selected pairs set $S_p \leftarrow \emptyset$.
    \ForAll{cell $h_k \in C_{\text{sel}}$}
        \State Randomly sample one pair of distinct observations $(s_a, s_b)$ from $P_k$.
        \State Add $(s_a, s_b)$ to $S_p$.
    \EndFor
    \State $q \leftarrow |S_p|$.
    \State \Return $S_p$.
\EndProcedure
\end{algorithmic}
\end{algorithm}
The pairs in $S_p$ are then used for PL estimation (Equations \ref{eq:pairwise_loglik_contrib_single}-\ref{eq:lambda_hat_pl_new}). Some examples of pair selections according with \ref{alg:sgpl} for different simulation scenarios are contained in the Appendix section \ref{sec:append}.

\subsection{Theoretical Justification}
The $k_{\text{ring}}$ isolation step is crucial. By ensuring that cells from which pairs are drawn are spatially separated, we aim to reduce the correlation between pairs sampled from different cells. While true independence is difficult to guarantee without knowing the exact range of the spatial process, this provides a practical mechanism to approximate the assumption that pairs $(s_a, s_b)$ from cell $h_k$ and $(s_c, s_d)$ from cell $h_m$ (where $h_k \neq h_m$) are stochastically independent, or at least have much weaker dependence than pairs within the same cell.

The grid overlay and cell-based sampling contribute significantly to computational efficiency. Assigning points to H3 cells is efficient. H3's indexing system allows for fast neighborhood queries (finding $k_{\text{ring}}$ neighbors). The most computationally intensive part of traditional methods, such as calculating all $N(N-1)/2$ pairwise distances or manipulating large $N \times N$ matrices, is avoided. The PL estimation itself is then performed on $q$ pairs, where $q$ is typically much smaller than $N$.

\subsection{Computational Complexity}
The computational advantages of SG-PL are substantial. Assigning $N$ points to H3 cells is typically $O(N)$ or $O(N \log N)$ depending on implementation. The iterative cell selection process depends on $q_{\text{target}}$ and the density of cells, but H3 neighborhood queries are very fast (constant time for a given $k_{\text{ring}}$). The dominant part of the SG-PL algorithm is the PL estimation on $q$ pairs, which involves iterative solutions to Equations \ref{eq:beta_hat_pl_new}-\ref{eq:lambda_hat_pl_new}. Each iteration involves sums over $q$ pairs and, in the general case of $p$ regressors, inversion of a $p \times p$ matrix, which is computationally trivial if $p$ is small. This contrasts sharply with the $O(N^3)$ or $O(N^2)$ complexities of traditional spatial econometric methods.

\section{Simulation Study}
\label{sec:simulation}
A Monte Carlo simulation study was conducted to assess the statistical accuracy and computational performance of the proposed SG-PL method. Its performance was compared against a standard benchmark, the Generalized Moments (GM) estimator for Spatial Error Models (SEM).

Data were generated from a Spatial Error Model (SEM):
\begin{equation}
    y_s = \beta_0 + \beta_1 x_{s,1} + u_s, \quad \text{for } s=1, \dots, N.
\end{equation}
The error vector $\vect{u} = (u_1, \dots, u_N)'$ was generated as:
\begin{equation}
    \vect{u} = (\mat{I}_N - \lambda_{\text{SEM}} \mat{W}_{\text{DGP}})^{-1} \vect{\epsilon},
\end{equation}
where $\vect{\epsilon} \sim N(\vect{0}, \sigma_{\epsilon}^2 \mat{I}_N)$, $\mat{I}_N$ is an $N \times N$ identity matrix, $\lambda_{\text{SEM}}$ is the true spatial autocorrelation parameter for the SEM, and $\mat{W}_{\text{DGP}}$ is the spatial weights matrix used in the data generation process. $\mat{W}_{\text{DGP}}$ was constructed using $k_{\text{W}_{\text{DGP}}}$ nearest neighbors, row-standardized.

For large $N$, direct computation of $(\mat{I}_N - \lambda_{\text{SEM}} \mat{W}_{\text{DGP}})^{-1}$ is $O(N^3)$. To make the DGP computationally feasible for large $N$, we approximated the term $(\mat{I}_N - \lambda_{\text{SEM}} \mat{W}_{\text{DGP}})^{-1}\vect{\epsilon}$ using a $K_{\text{Taylor}}$-term Taylor/Neumann series expansion:
\begin{equation}
    \vect{u} \approx \sum_{k=0}^{K_{\text{Taylor}}} (\lambda_{\text{SEM}} \mat{W}_{\text{DGP}})^k \vect{\epsilon}.
\end{equation}
This approximation is common in simulations of spatial models for large $N$ (e.g., \citet{LeSage2009}).

Spatial point patterns for the $N$ observations were generated under two scenarios. The first scenario, termed "Uniform," involved sampling $N$ points uniformly at random within a unit square $[0,1] \times [0,1]$. The second scenario, "Clustered," began by uniformly distributing $N_c \sim \text{Poisson}(\lambda_c)$ centroids in the unit square. Subsequently, approximately $N/N_c$ points were generated around each centroid $\mu_j$ using a bivariate Normal distribution $N(\mu_j, \mat{\Sigma}_{\text{cluster}})$ with $\mat{\Sigma}_{\text{cluster}} = \text{diag}(\sigma_{\text{cluster}}^2, \sigma_{\text{cluster}}^2)$, ensuring points remained within the unit square by rejection sampling. The single explanatory variable $x_{s,1}$ was drawn from $N(\mu_x, \sigma_x^2)$.

Fixed true parameter values for the DGP were set as follows: $\beta_0 = 1.0$ and $\beta_1 = 1.5$. The explanatory variable $x_{s,1}$ was drawn with $\mu_x = 0$ and $\sigma_x = 1$. For clustered data, the expected number of clusters $\lambda_c$ was 5, and the dispersion around cluster centers $\sigma_{\text{cluster}}$ was 0.05. The true error variance for $\epsilon$, $\sigma_{\epsilon}^2$, was 0.1.

The study compared two methods. The first was the Sampled Grid Pairwise Likelihood (SG-PL) method, as described in Section \ref{sec:sgpl_method}. For SG-PL in the simulation, the H3 resolution $R_{H3}$ was set to 7, the k-ring separation $k_{\text{ring}}$ to 1, the target number of pairs $q_{\text{target}}$ to 1000, and the minimum points per H3 cell $N_{\min,H3}$ to 2.
The second method, serving as a benchmark, was the Generalized Moments (GM) SEM Estimator. This was implemented using the `GMerrorsar` function from the `spatialreg` R package \citep{spatialregpackage} to estimate the SEM on the full dataset. The spatial weights matrix for GM estimation, $\mat{W}_{\text{GM}}$, was constructed using $k_{\text{W}_{\text{GM}}}=4$ nearest neighbors. This is a well-established and computationally feasible (though still more intensive than PL) method for SEM estimation on moderately large datasets.

\subsection{Simulation Design}
In the experimentation we varying several parameters: the total number of points $N$ took values from the set $\{5000, 10000, 25000\}$; the true SEM spatial autocorrelation $\lambda_{\text{SEM}}$ was chosen from $\{-0.3, 0, 0.3, 0.7\}$; and the spatial point pattern type was either "uniform" or "clustered".
For each scenario, representing a unique combination of $N$, $\lambda_{\text{SEM}}$, and data type, one dataset was generated. On this fixed dataset, $M_{\text{reps}}=500$ runs of the SG-PL method (including the random cell selection and pair sampling steps, followed by PL estimation) were performed. This allows assessment of the variability of SG-PL estimates due to its inherent sampling randomness. The GM-SEM estimator was run once per generated dataset as it is deterministic given the data and $\mat{W}_{\text{GM}}$.

The performance of the estimators for $\beta_1$, the spatial parameter ($\lambda_{\text{PL}}$ for PL, $\lambda_{\text{GM}}$ for GM), and the error variance ($\sigma^2_{\text{PL}}$ for PL, $\sigma_{\epsilon}^2$ for GM) was assessed using several metrics. These included the mean computational time (in seconds) for one estimation run of SG-PL and GM. Another key metric was Relative Efficiency (RE), calculated as $RE(\text{SG-PL vs GM for } \theta) = \frac{\text{MSE}(\hat{\theta}_{\text{GM}})}{\text{MSE}(\hat{\theta}_{\text{SG-PL}})}$. An RE value greater than 1 indicates that SG-PL is more statistically efficient (has a lower MSE) than GM for parameter $\theta$, while an RE value less than 1 indicates GM is more efficient. For SG-PL, Bias and MSE were calculated using the mean of the estimates over the $M_{\text{reps}}=500$ runs.

\subsection{Simulation Results and Discussion}

\begin{figure}[htbp]
    \centering
    \includegraphics[width=1\textwidth, keepaspectratio]{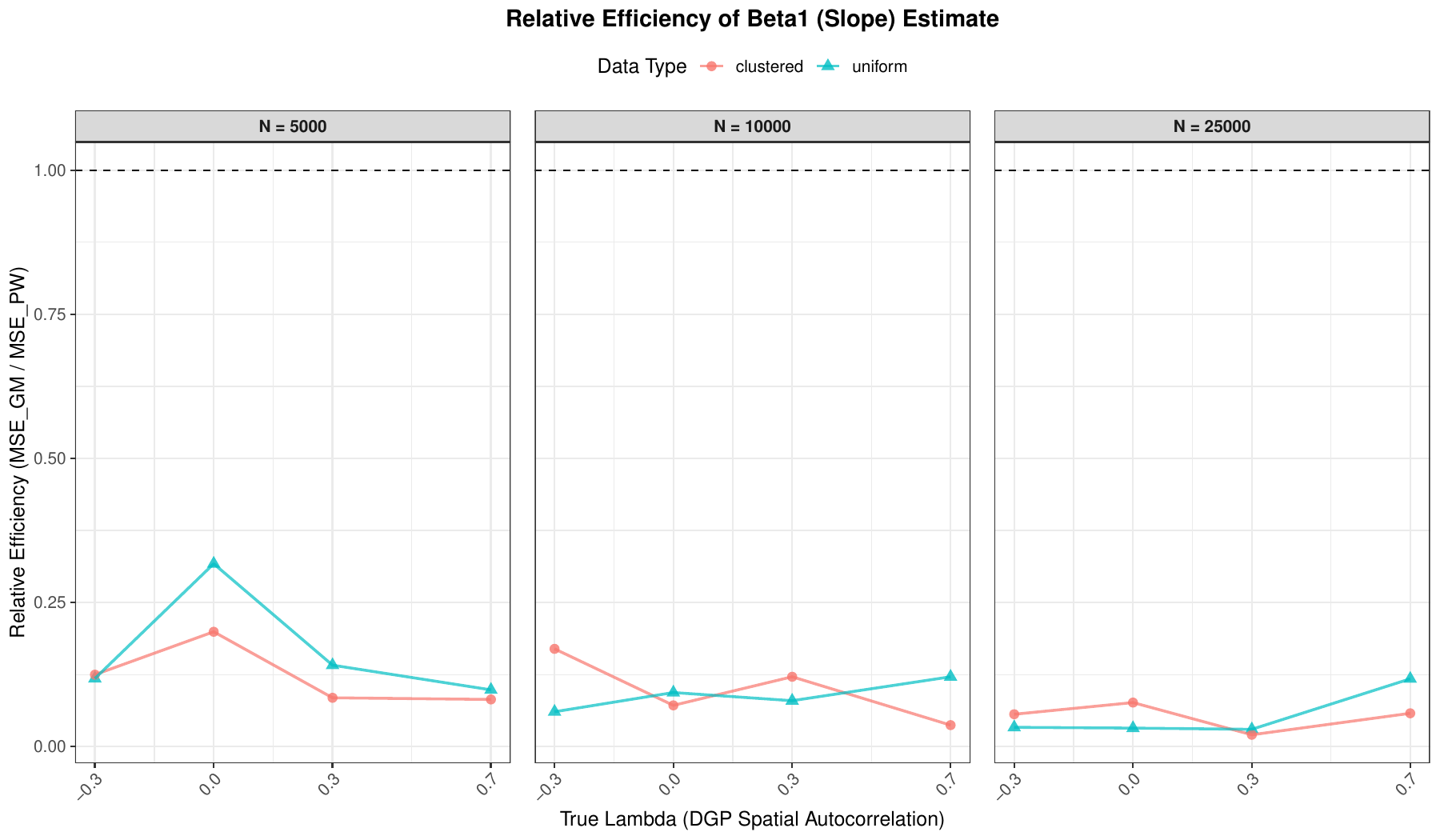}
    \caption{Relative Efficiency of $\hat{\beta}_1$ (Slope) Estimate (SG-PL vs GM). Values below the dashed line (RE=1) indicate GM is more statistically efficient.}
    \label{fig:re_beta}
\end{figure}

\begin{figure}[htbp]
    \centering
    \includegraphics[width=1\textwidth, keepaspectratio]{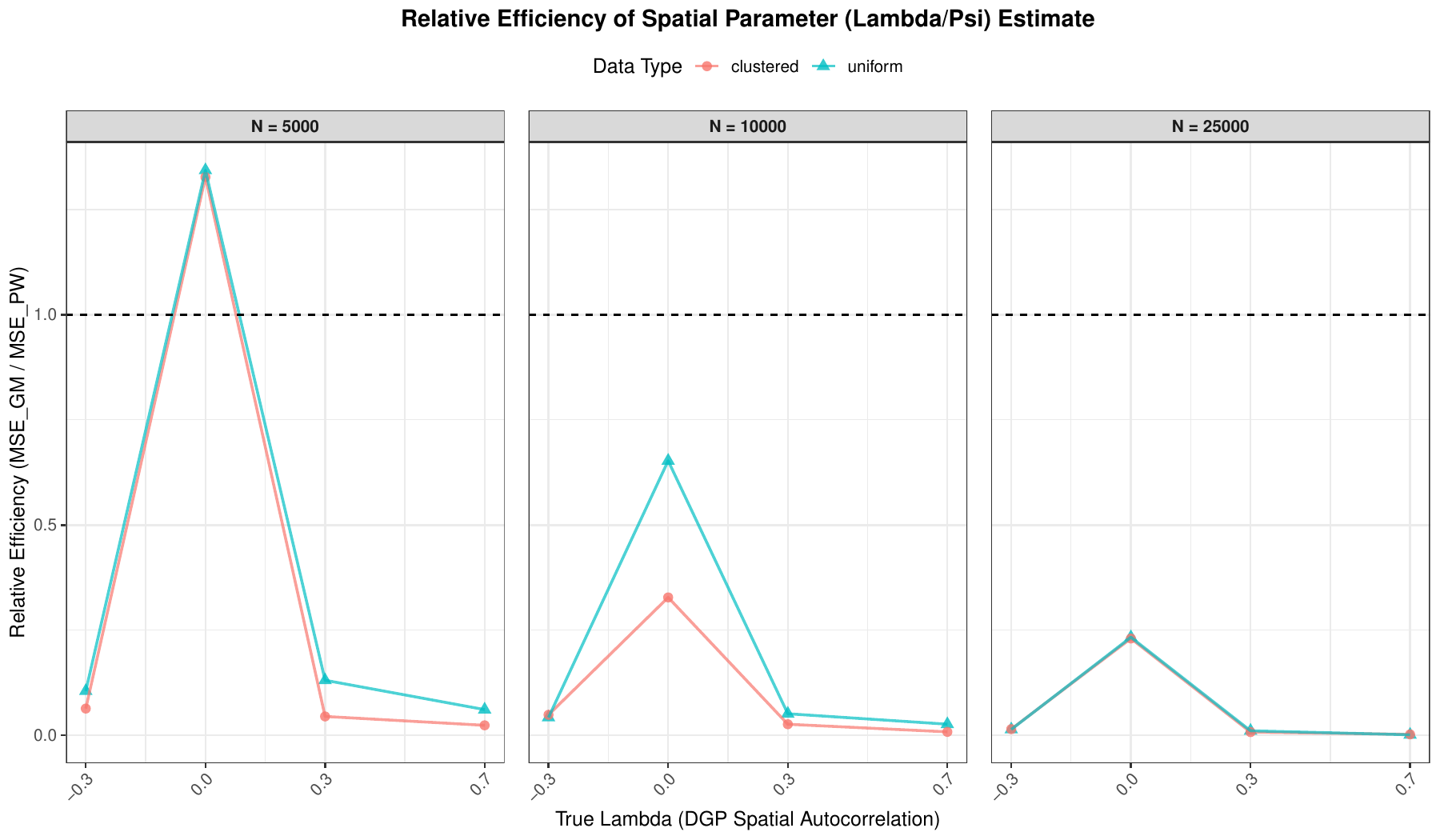}
    \caption{Relative Efficiency of Spatial Parameter ($\hat{\lambda}_{\text{PL}}$ vs $\hat{\lambda}_{\text{GM}}$) Estimate. Values above RE=1 indicate SG-PL is more efficient.}
    \label{fig:re_spatial_param}
\end{figure}

Tables \ref{tab:re_uniform} and \ref{tab:re_clustered} present the Relative Efficiency (RE) of SG-PL compared to GM-SEM for the slope coefficient ($\hat{\beta}_1$), the spatial autocorrelation parameter ($\hat{\lambda}_{\text{PL}}$ vs $\hat{\lambda}_{\text{GM}}$), and the error variance ($\hat{\sigma}^2_{\text{PL}}$ vs $\hat{\sigma}^2_{\epsilon,\text{GM}}$) under uniform and clustered data scenarios, respectively. Figures \ref{fig:re_beta}, \ref{fig:re_spatial_param}, and \ref{fig:re_sigma_sq} visualize these RE values.

\begin{table}[htbp]
\centering
\caption{Relative Efficiency (RE) of SG-PL vs. GM - Uniform Data Scenarios}
\label{tab:re_uniform}
\begin{tabular}{lrrrrr}
\toprule
$N$ & {$\lambda_{SEM}$} & {$RE(\hat{\beta}_{1})$} & {$RE(\text{spatial param.})$} & {$RE(\hat{\sigma}^2)$} \\
\midrule
5000  & -0.3 & 0.118  & 0.105  & 6.78e-5 \\
10000 & -0.3 & 0.0603 & 0.0426 & 0.0115 \\
25000 & -0.3 & 0.0334 & 0.0140 & 0.00353 \\
\midrule
5000  & 0.0  & 0.317  & 1.34   & 0.0106 \\
10000 & 0.0  & 0.0937 & 0.653  & 0.00529 \\
25000 & 0.0  & 0.0320 & 0.233  & 0.0589 \\
\midrule
5000  & 0.3  & 0.141  & 0.131  & 3.28e-5 \\
10000 & 0.3  & 0.0796 & 0.0513 & 0.00168 \\
25000 & 0.3  & 0.0299 & 0.0108 & 0.00341 \\
\midrule
5000  & 0.7  & 0.0984 & 0.0611 & 1.69e-4 \\
10000 & 0.7  & 0.121  & 0.0264 & 4.69e-5 \\
25000 & 0.7  & 0.117  & 0.00127& 4.05e-4 \\
\bottomrule
\end{tabular}
\end{table}

\begin{table}[htbp]
\centering
\caption{Relative Efficiency (RE) of SG-PL vs. GM - Clustered Data Scenarios}
\label{tab:re_clustered}
\begin{tabular}{lrrrr}
\toprule
$N$ & {$\lambda_{SEM}$} & {$RE(\hat{\beta}_{1})$} & {$RE(\text{spatial param.})$} & {$RE(\hat{\sigma}^2)$} \\
\midrule
5000  & -0.3 & 0.125  & 0.0636 & 0.00287 \\
10000 & -0.3 & 0.170  & 0.0488 & 0.0107 \\
25000 & -0.3 & 0.0560 & 0.0147 & 0.00989 \\
\midrule
5000  & 0.0  & 0.199  & 1.33   & 6.41e-5 \\
10000 & 0.0  & 0.0715 & 0.328  & 3.94e-4 \\
25000 & 0.0  & 0.0763 & 0.230  & 0.0396 \\
\midrule
5000  & 0.3  & 0.0846 & 0.0449 & 0.0664 \\
10000 & 0.3  & 0.121  & 0.0263 & 0.0928 \\
25000 & 0.3  & 0.0204 & 0.00767& 0.00592 \\
\midrule
5000  & 0.7  & 0.0817 & 0.0238 & 8.98e-4 \\
10000 & 0.7  & 0.0371 & 0.00821& 5.53e-4 \\
25000 & 0.7  & 0.0577 & 0.00224& 2.23e-4 \\
\bottomrule
\end{tabular}
\end{table}

\begin{figure}[htbp]
    \centering
    \includegraphics[width=1\textwidth, keepaspectratio]{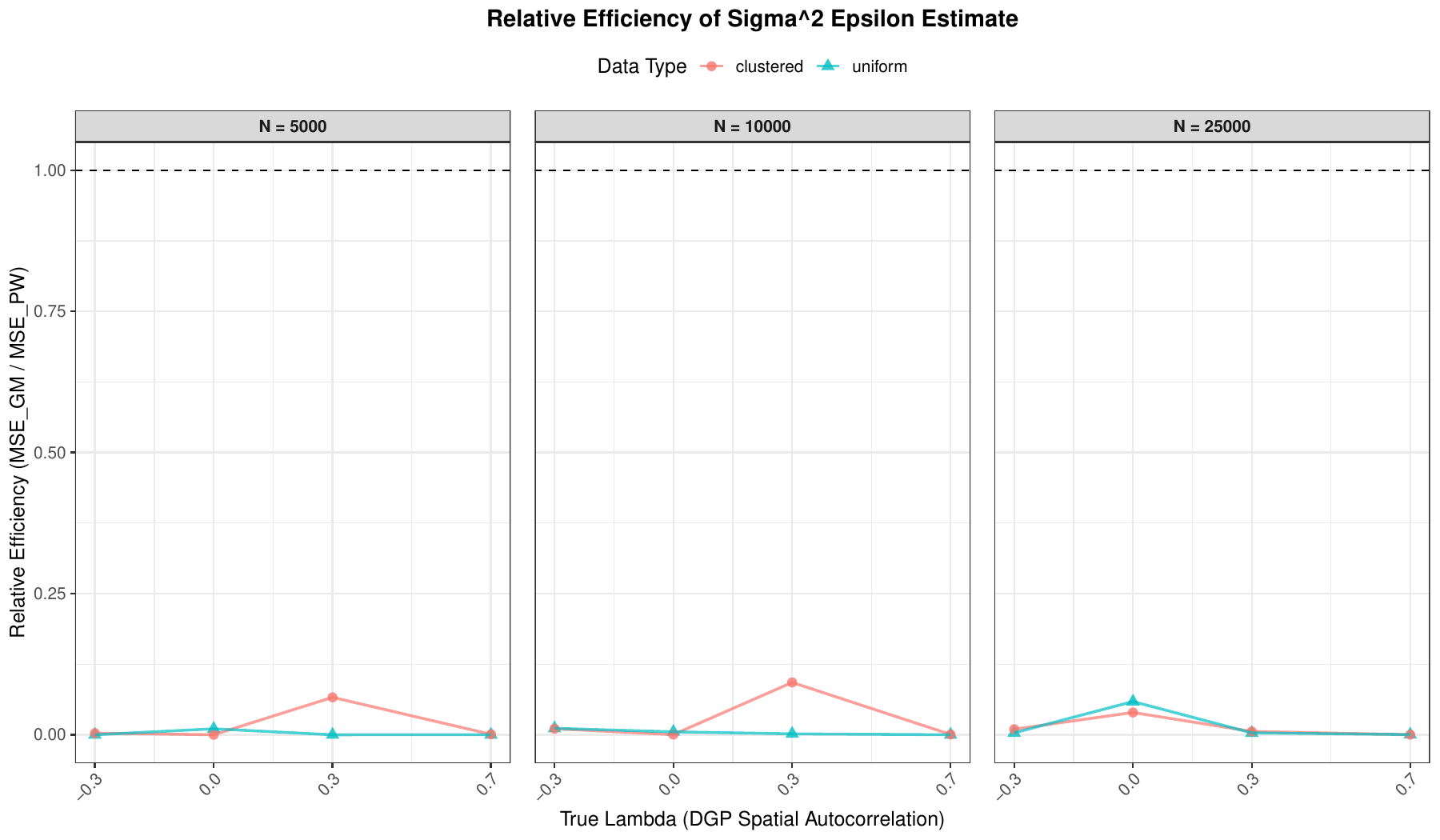}
    \caption{Relative Efficiency of $\hat{\sigma}^2_{\epsilon}$ (Error Variance) Estimate (SG-PL vs GM). Values below RE=1 indicate GM is more statistically efficient.}
    \label{fig:re_sigma_sq}
\end{figure}

For the slope coefficient $\hat{\beta}_1$ (Figure \ref{fig:re_beta}), the RE values are consistently below 1 across almost all scenarios for both uniform and clustered data. This indicates that the GM-SEM estimator is generally more statistically efficient (lower MSE) for $\hat{\beta}_1$ than SG-PL. The RE for SG-PL tends to decrease (indicating worsening relative statistical efficiency) as the sample size $N$ increases, particularly when the true spatial autocorrelation $\lambda_{\text{SEM}}$ is non-zero. For instance, with uniform data and $\lambda_{\text{SEM}}=-0.3$, RE for $\hat{\beta}_1$ drops from 0.118 at $N=5000$ to 0.0334 at $N=25000$. This pattern suggests that as more information becomes available with larger $N$, the full-information nature of GM (even approximated) allows it to achieve better precision for $\beta_1$ compared to the limited-pair approach of SG-PL (that is estimate with the same number of pairs).

Regarding the spatial parameter (Figure \ref{fig:re_spatial_param}, comparing $\hat{\lambda}_{\text{PL}}$ from SG-PL and $\hat{\lambda}_{\text{GM}}$ from GM-SEM), the results are more nuanced. Notably, when the true spatial autocorrelation $\lambda_{\text{SEM}}=0$ and $N=5000$, SG-PL exhibits an RE greater than 1 (1.34 for uniform, 1.33 for clustered data). This suggests that under conditions of no true spatial dependence and smaller sample sizes in this study, SG-PL can be more statistically efficient than GM-SEM in estimating the (absence of) spatial correlation. This is a significant finding, as accurately detecting weak or zero spatial effects is important. However, for stronger true spatial correlation ($\lambda_{\text{SEM}} \neq 0$) and for larger $N$ even when $\lambda_{\text{SEM}}=0$, GM-SEM generally shows better statistical efficiency (RE $<$ 1). For example, with uniform data and $\lambda_{\text{SEM}}=0.7$, RE for the spatial parameter drops from 0.0611 at $N=5000$ to a very low 0.00127 at $N=25000$.

For the error variance component $\hat{\sigma}^2$ (Figure \ref{fig:re_sigma_sq}), the RE values for SG-PL are consistently very low across all scenarios (often several orders of magnitude less than 1). This indicates that the GM-SEM method is substantially more statistically efficient for estimating the error variance $\sigma^2_{\epsilon}$. This is not entirely surprising, as the $\sigma^2_{\text{PL}}$ in the pairwise likelihood (Equation \ref{eq:bvn_density_errors}) is the marginal variance of the bivariate error distribution, which is related but not identical to $\sigma^2_{\epsilon}$ in the SEM, and the PL approach uses far less information to estimate it.

In general, the lower statistical efficiency of SG-PL for most parameters is an expected consequence of using only a small subset ($q_{\text{target}}=1000$ pairs) of the available information compared to the GM-SEM method, which utilizes information from the entire dataset to construct its moment conditions. Pairwise likelihood itself is an approximation to the full likelihood, and further sampling of pairs can improve the  efficiency.

\begin{figure}[htbp]
    \centering
    \includegraphics[width=1\textwidth, keepaspectratio]{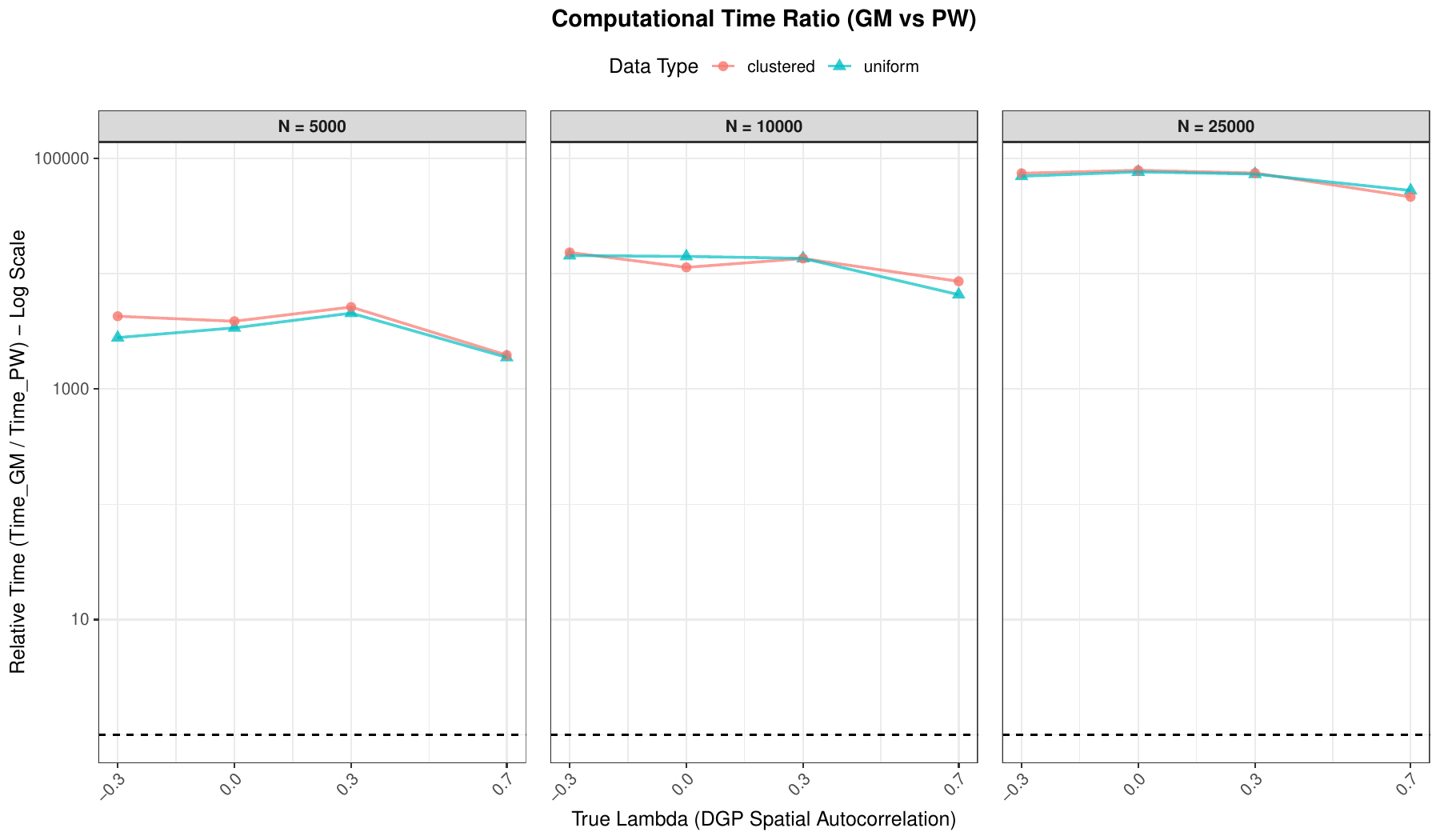}
    \caption{Relative Computational Time (GM Time / SG-PL Time) on a Logarithmic Scale. Higher values indicate SG-PL is faster.}
    \label{fig:rel_time}
\end{figure}

Tables \ref{tab:time_uniform} and \ref{tab:time_clustered} present the mean computational times for SG-PL and GM-SEM. Figure \ref{fig:rel_time} visualizes the relative computational time (GM time / SG-PL time) on a logarithmic scale.

\begin{table}[htbp]
\centering
\caption{Computational Time Comparison (Seconds) - Uniform Data Scenarios}
\label{tab:time_uniform}
\begin{tabular}{lrrrrr}
\toprule
$N$ & {$\lambda_{SEM}$} & {Mean Time (GM)} & {Mean Time (SG-PL)} & {Relative Time (GM / SG-PL)} \\
\midrule
5000  & -0.3 & 0.967 & 3.47e-4 & 2.79e3 \\
10000 & -0.3 & 2.76  & 1.93e-4 & 1.43e4 \\
25000 & -0.3 & 14.7  & 2.10e-4 & 6.99e4 \\
\midrule
5000  & 0.0  & 0.859 & 2.54e-4 & 3.38e3 \\
10000 & 0.0  & 2.91  & 2.06e-4 & 1.41e4 \\
25000 & 0.0  & 14.5  & 1.90e-4 & 7.62e4 \\
\midrule
5000  & 0.3  & 0.943 & 2.08e-4 & 4.54e3 \\
10000 & 0.3  & 2.73  & 2.02e-4 & 1.36e4 \\
25000 & 0.3  & 14.2  & 1.95e-4 & 7.29e4 \\
\midrule
5000  & 0.7  & 0.889 & 4.74e-4 & 1.88e3 \\
10000 & 0.7  & 2.76  & 4.21e-4 & 6.57e3 \\
25000 & 0.7  & 14.1  & 2.69e-4 & 5.26e4 \\
\bottomrule
\end{tabular}
\end{table}

\begin{table}[htbp]
\centering
\caption{Computational Time Comparison (Seconds) - Clustered Data Scenarios}
\label{tab:time_clustered}
\begin{tabular}{lrrrrr}
\toprule
$N$ & {$\lambda_{SEM}$} & {Mean Time (GM)} & {Mean Time (SG-PL)} & {Relative Time (GM / SG-PL)} \\
\midrule
5000  & -0.3 & 0.858 & 2.01e-4 & 4.27e3 \\
10000 & -0.3 & 2.88  & 1.88e-4 & 1.53e4 \\
25000 & -0.3 & 14.5  & 1.96e-4 & 7.42e4 \\
\midrule
5000  & 0.0  & 0.912 & 2.36e-4 & 3.86e3 \\
10000 & 0.0  & 2.74  & 2.42e-4 & 1.13e4 \\
25000 & 0.0  & 16.8  & 2.13e-4 & 7.88e4 \\
\midrule
5000  & 0.3  & 1.19  & 2.32e-4 & 5.13e3 \\
10000 & 0.3  & 3.24  & 2.40e-4 & 1.35e4 \\
25000 & 0.3  & 15.1  & 2.02e-4 & 7.45e4 \\
\midrule
5000  & 0.7  & 0.827 & 4.22e-4 & 1.96e3 \\
10000 & 0.7  & 2.95  & 3.44e-4 & 8.57e3 \\
25000 & 0.7  & 14.4  & 3.10e-4 & 4.63e4 \\
\bottomrule
\end{tabular}
\end{table}

The SG-PL method demonstrates a dramatic and crucial advantage in computational time. As seen in Tables \ref{tab:time_uniform}, \ref{tab:time_clustered}, and strikingly in Figure \ref{fig:rel_time}, SG-PL is consistently orders of magnitude faster than the GM-SEM benchmark. The relative speed-up (GM Time / SG-PL Time) ranges from approximately 1,880 times faster (for $N=5000, \lambda_{\text{SEM}}=0.7$, uniform) to nearly 79,000 times faster (for $N=25000, \lambda_{\text{SEM}}=0.0$, clustered). This computational advantage generally becomes more pronounced as the sample size $N$ increases, underscoring SG-PL's scalability. For $N=25000$, GM-SEM takes around 14-17 seconds, while SG-PL takes only a few ten-thousandths of a second (i.e., fractions of a millisecond) for its core estimation once pairs are selected. The pair selection itself is also very fast. This is the primary strength of SG-PL, making it a viable tool for datasets where GM or full Maximum Likelihood methods are computationally infeasible due to their $O(N^2)$ or $O(N^3)$ complexities.

The simulation results clearly highlight a critical trade-off: SG-PL achieves substantial computational gains at the cost of some statistical efficiency for most parameters, particularly $\beta_1$ and $\sigma^2$. This trade-off is often acceptable or indeed necessary when dealing with very large datasets where more statistically efficient methods are simply too slow to run. The ability to obtain estimates rapidly makes SG-PL a valuable tool for exploratory data analysis, for applications with severe time constraints, or when computational resources are limited.

Overall, the simulation results underscore that the SG-PL method's primary contribution is its significant computational speed-up, which makes pairwise likelihood estimation practical for spatial regression on very large, irregular datasets.

\section{Empirical Application: King County House Sales}
\label{sec:empirical}
To demonstrate the applicability and performance of SG-PL on real-world data, we analyzed a large dataset of house sales in King County, Washington, USA, comparing our method against two established benchmarks: the Generalized Moments (GM) estimator and the full Maximum Likelihood (ML) estimator.

The dataset comprises $N=21,613$ house sales recorded between May 2014 and May 2015. For this application, we estimate a simple hedonic price model where the sales price (`price`) is the dependent variable, and the square footage of living area (`sqft\_liv`) is the primary independent variable. The spatial coordinates (latitude and longitude) of each house are used to define the spatial relationships.

We applied the SG-PL method to estimate a spatial error model of the form:
\begin{equation}
    \text{price}_s = \beta_0 + \beta_1 \text{sqft\_liv}_s + u_s,
\end{equation}
where $u_s$ are errors assumed to follow the pairwise spatial structure outlined in Section \ref{sec:background}. For the SG-PL estimation, we used a fine-grained H3 grid with resolution $R_{H3}=9$. The results were averaged over 100 runs of the random cell selection and pair sampling procedure to ensure stable estimates. The intercept $\beta_0$ was concentrated out by demeaning the variables based on full sample means before applying the PL procedure.

As benchmarks, we estimated the same SEM model using two methods from the `spatialreg` R package on the entire dataset. The first is the full Maximum Likelihood (ML) estimator via the `errorsarlm` function, which serves as the gold standard for statistical efficiency but is computationally intensive. The second is the Generalized Moments (GM) estimator via the `GMerrorsar` function, which is a common alternative for large datasets. For both benchmarks, the spatial weights matrix $\mat{W}$ was constructed using $k=5$ nearest neighbors.

Table \ref{tab:kingcounty_results_revised} presents the parameter estimates from the SG-PL method (averaged over 100 runs) alongside the results from the ML and GM benchmark estimators.

\begin{table}[htbp]
\centering
\caption{Comparison of Parameter Estimates for King County House Sales Data}
\label{tab:kingcounty_results_revised}
\begin{tabular}{lrrr} 
\toprule
{Parameter} & {SG-PL (avg. of 100 runs)} & {ML-SEM (Benchmark)} & {GM-SEM (Benchmark)} \\
\midrule
$\hat{\beta}_1$ (sqft\_liv) & 247.35 & 228.11 & 255.58 \\ 
$\hat{\lambda}$ & 0.720 & 0.748 & 0.358 \\
$\hat{\sigma}^2$ & \num{6.77e10} & \num{3.18e10} & \num{4.56e10} \\ 
\bottomrule
\end{tabular}
\end{table}

The empirical results compellingly demonstrate the strength of the SG-PL method. In real-world applications where true parameters are unknown, the goal is to balance statistical accuracy with computational feasibility. The SG-PL method excels in this trade-off, providing robust estimates with unparalleled speed.

The estimated slope coefficient, $\hat{\beta}_1$, from SG-PL is 247.35. This value is highly plausible, falling between the ML estimate (228.11) and the GM estimate (255.58). This shows that SG-PL effectively captures the core economic relationship between living area and house price, yielding a coefficient that is consistent with established, computationally intensive methods.

The most striking result is found in the estimation of the spatial parameter, $\hat{\lambda}$. The SG-PL estimate of $\hat{\lambda}_{\text{PL}} = 0.720$ is remarkably close to the gold-standard ML estimate of $\hat{\lambda}_{\text{ML}} = 0.748$. In contrast, the GM method substantially underestimates the strength of the spatial dependence, yielding a much lower $\hat{\lambda}_{\text{GM}} = 0.358$. This is a critical finding: despite its simplicity and speed, SG-PL provides a far more accurate estimate of the underlying spatial structure than the more complex GM estimator. This suggests that the localized, isolated pairing strategy of SG-PL is highly effective at capturing the true extent of spatial correlation in the data-generating process.

Regarding the error variance, the estimates differ across the three methods. As established in our simulation study, the pairwise variance $\hat{\sigma}^2_{\text{PL}}$ is a theoretically different quantity from the error variance ($\hat{\sigma}^2_{\epsilon}$) estimated by SEM models and is known to be estimated with lower statistical efficiency. This is an expected and acceptable trade-off for the substantial gains in other areas.

In conclusion, this empirical application highlights the exceptional practical value of SG-PL. It not only produces key regression coefficients that are in line with benchmarks but, more importantly, it delivers a significantly more accurate estimate of the spatial correlation parameter than the GM method, rivaling the precision of the full ML estimator. This superior accuracy, combined with the orders-of-magnitude reduction in computational time documented in our simulations, establishes SG-PL as a powerful, scalable, and reliable tool for rapid and accurate exploratory analysis and modeling of very large spatial datasets.

\section{Conclusion and Future Work}
\label{sec:conclusion}

This paper introduced the Sampled Grid Pairwise Likelihood (SG-PL) method, a novel and computationally highly efficient approach for applying pairwise likelihood estimation to spatial regression models on large, irregular spatial datasets. SG-PL uniquely leverages a regular discrete global grid system (H3) overlay for structured and spatially isolated sampling of observation pairs. This design directly addresses the computational bottleneck of pair selection and estimation in traditional PL applications to large $N$.

The simulation study unequivocally demonstrated SG-PL's primary strength: it can be orders of magnitude faster than a benchmark Generalized Moments (GM) SEM estimator. This computational advantage is not static; crucially, the relative speed-up of SG-PL increases with the sample size (N), rendering it an exceptionally practical tool for datasets that are too large for conventional methods. This remarkable computational efficiency is accompanied by a trade-off in statistical efficiency. While the GM-SEM estimator generally showed lower MSE for some parameters like regression coefficients and error variance, the loss in statistical efficiency with SG-PL is considerably outweighed by the massive gains in computational speed. This balance shifts even more favorably towards SG-PL as N grows, because the computational time savings become increasingly vital while the efficiency cost remains comparatively modest. SG-PL also exhibited promising statistical efficiency for the spatial correlation parameter in specific scenarios, such as when the true spatial correlation was zero. The empirical application further underscored SG-PL's practical value, delivering comparable key estimates to the other estimation methods at a dramatically reduced computational cost.

Future research could focus on extending the SG-PL framework in several directions. A natural generalization is to move beyond pairs to sample triplets or other larger subsets of observations within the selected grid cells, evolving the method from a Pairwise Likelihood to a more general Composite Likelihood approach. This may offer improvements in statistical efficiency. Furthermore, adapting the SG sampling strategy to other complex models, such as Spatial Autoregressive (SAR) models, Spatial Durbin Models (SDM), spatial panel data models, or models for non-Gaussian spatial data (e.g., for binary or count data), would significantly broaden its applicability.

In conclusion, the SG-PL method offers a significant addition to the spatial statistician's toolkit. It provides a highly scalable and computationally feasible approach for analyzing large, irregular spatial datasets, primarily due to its dramatic, orders-of-magnitude speed advantages over traditional methods. While this computational prowess involves a trade-off with statistical efficiency, the magnitude of the speed gain far surpasses the reduction in estimator precision. Importantly, this favorable trade-off becomes even more compelling as dataset sizes (N) increase, with the computational benefits of SG-PL growing substantially. This makes SG-PL particularly attractive for exploratory analysis and applications involving very large datasets where computational resources or time are primary constraints, offering a practical and increasingly advantageous balance between computational feasibility and reasonable statistical inference.

\newpage
\section{Appendix}
\label{sec:append}
\subsection{Code availability}
Full codes of simulation are available on Gihtub at \url{https://github.com/vincnardelli/pairwise-h3}. 

\subsection{Simulation output}
\begin{figure}[htbp]
    \centering
    \includegraphics[width=1\textwidth, keepaspectratio]{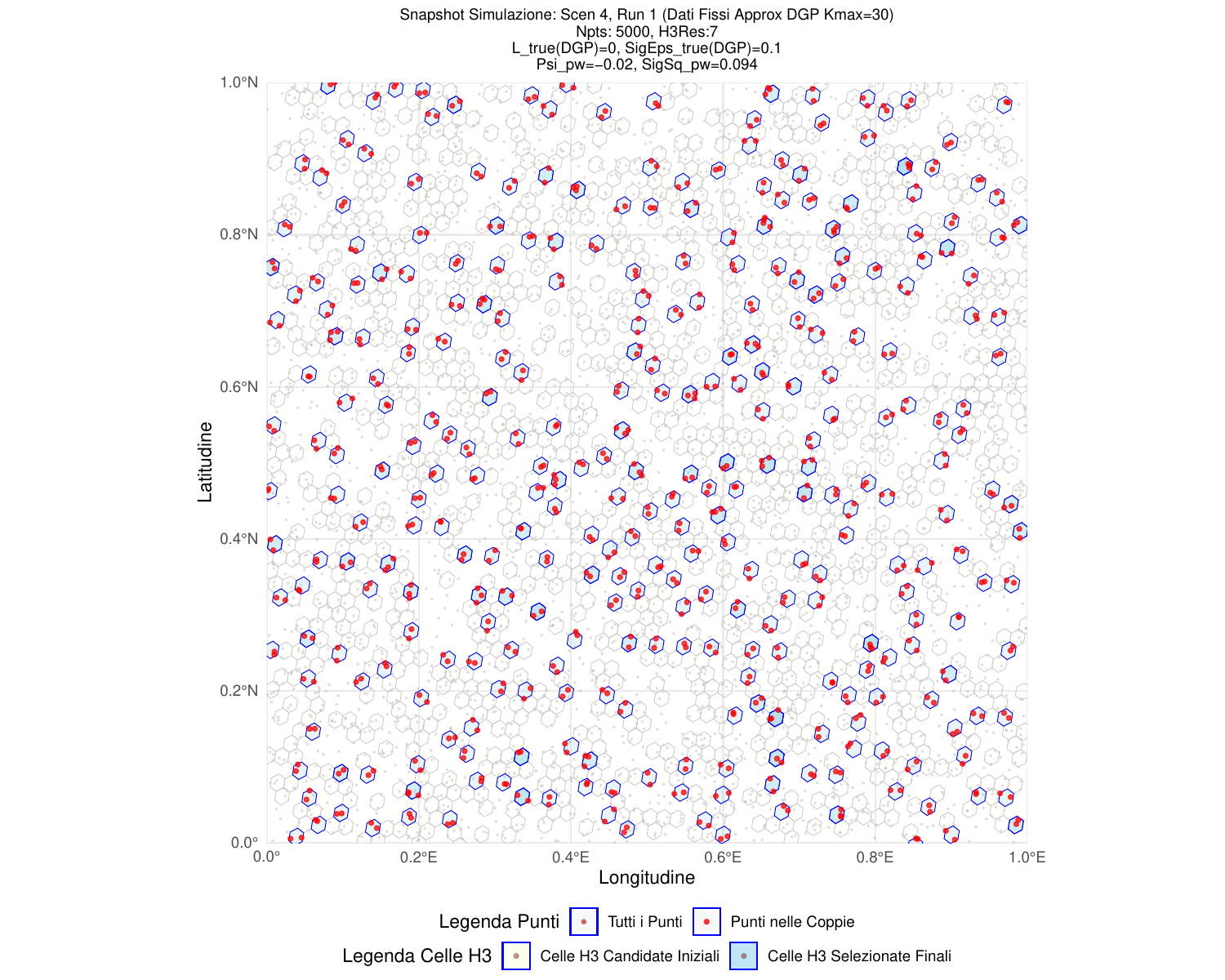}
    \caption{Observation pairs selected for $n = 10000$ for H3 resolution 7, data is uniform where $\lambda = 0.3$ and $\epsilon = 0.1$}
    \label{fig:re_sigma_sq}
\end{figure}

\begin{figure}[htbp]
    \centering
    \includegraphics[width=1\textwidth, keepaspectratio]{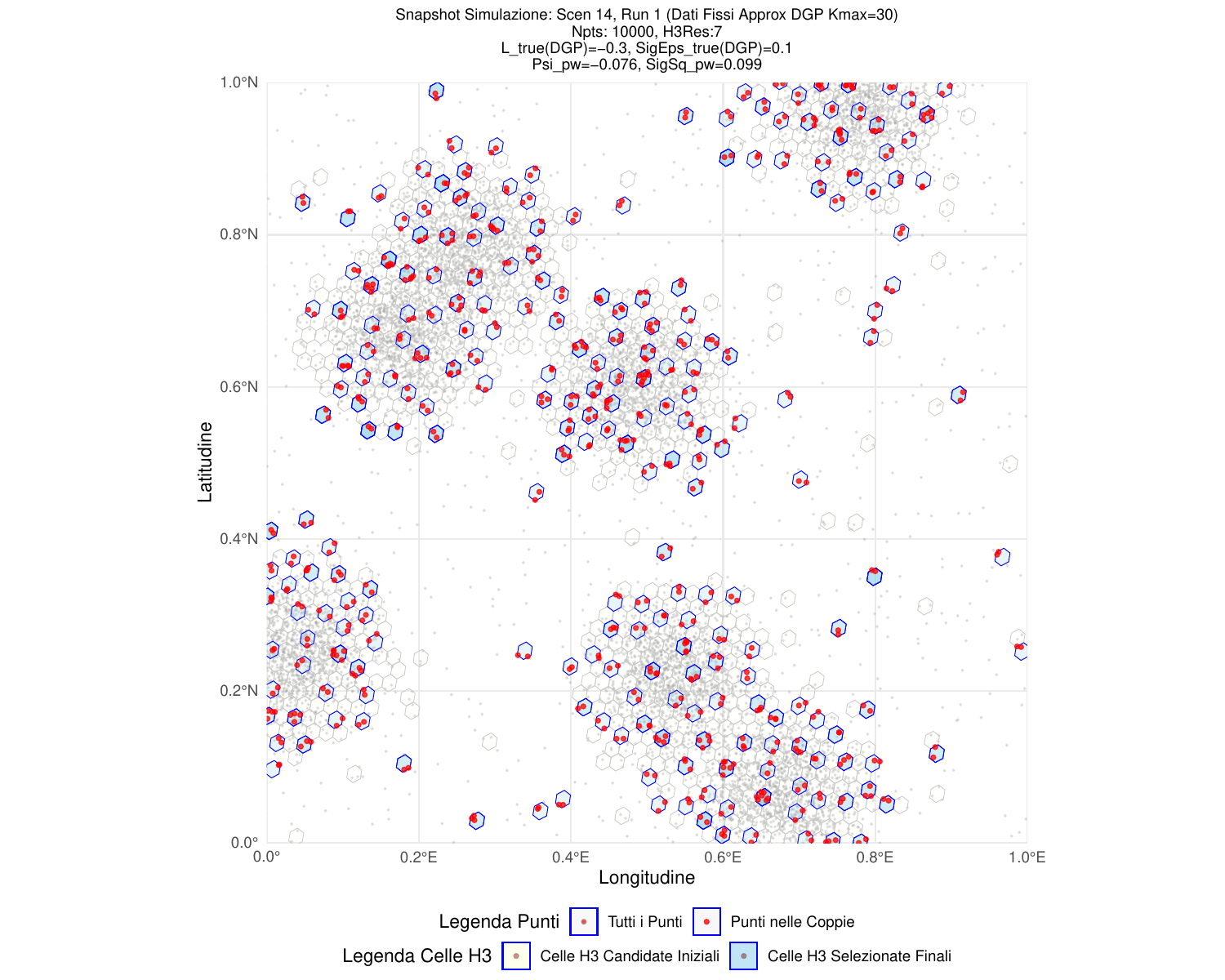}
    \caption{Observation pairs selected for $n = 10000$ for H3 resolution 7, data is uniform where $\lambda = 0.3$ and $\epsilon = 0.1$}
    \label{fig:re_sigma_sq}
\end{figure}

\begin{figure}[htbp]
    \centering
    \includegraphics[width=1\textwidth, keepaspectratio]{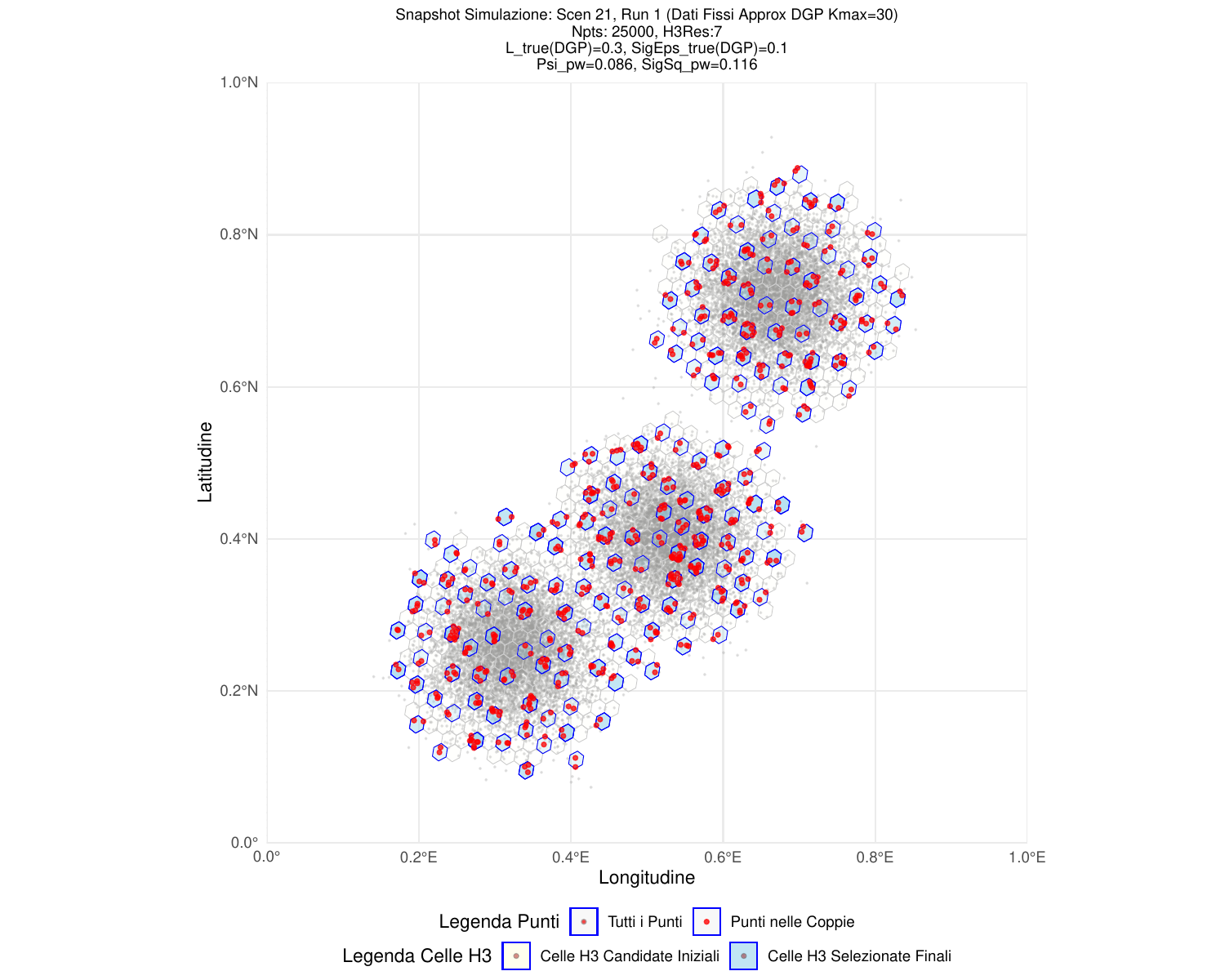}
    \caption{Observation pairs selected for $n = 25000$ for H3 resolution 7, data is clustered with 3  where $\lambda = 0.3$ and $\epsilon = 0.1$ }
    \label{fig:re_sigma_sq}
\end{figure}

\newpage
\bibliography{bibliography} 

\begin{thebibliography}{}

\bibitem[\protect\citeauthoryear{Arbia}{Arbia}{2014}]{Arbia2014}
Arbia, G. (2014).
\newblock Pairwise likelihood inference for spatial regressions estimated on very large datasets.
\newblock {\em Spatial Statistics\/}~{\em 7}, 21--39.

\bibitem[\protect\citeauthoryear{Arbia and Salvini}{Arbia and Salvini}{2024}]{ArbiaSalvini2024}
Arbia, G. and N.~Salvini (2024).
\newblock Feasible pairwise pseudo-likelihood inference on spatial regressions in large and irregular grids.
\newblock {\em ArXiv preprint arXiv:2401.05905\/}.

\bibitem[\protect\citeauthoryear{Bivand, Piras, Anselin, et~al.}{Bivand et~al.}{2013}]{spatialregpackage}
Bivand, R.~S., G.~Piras, L.~Anselin, et~al. (2013).
\newblock {\em spatialreg: Spatial Regression Analysis}.
\newblock R package version 1.2-8 (or current version).

\bibitem[\protect\citeauthoryear{LeSage and Pace}{LeSage and Pace}{2009}]{LeSage2009}
LeSage, J.~P. and R.~K. Pace (2009).
\newblock {\em Introduction to Spatial Econometrics}.
\newblock CRC Press.

\bibitem[\protect\citeauthoryear{Lindsay}{Lindsay}{1988}]{Lindsay1988}
Lindsay, B.~G. (1988).
\newblock Composite likelihood methods.
\newblock {\em Contemporary Mathematics\/}~{\em 80}, 221--239.

\bibitem[\protect\citeauthoryear{{Uber Technologies Inc.}}{{Uber Technologies Inc.}}{Nd}]{UberH3}
{Uber Technologies Inc.} (N.d.).
\newblock H3: Uber's hexagonal hierarchical spatial index.
\newblock Accessed 2025.

\bibitem[\protect\citeauthoryear{Varin, Reid, and Firth}{Varin et~al.}{2011}]{Varin2011}
Varin, C., N.~Reid, and D.~Firth (2011).
\newblock An overview of composite likelihood methods.
\newblock {\em Statistica Sinica\/}~{\em 21\/}(1), 5--42.

\end{thebibliography}

\end{document}